\documentstyle[11pt,twoside,jltp,epsfig]{article}

\title{Systematic Evolution of the 
Magnetotransport Properties of Bi$_{2}$Sr$_{2-x}$La$_x$CuO$_{6}$ 
with \\
Carrier Concentration}

\author{Yoichi Ando,$^{1,2 }$\address{$^{1}$Central Research Institute of 
Electric Power Industry, Tokyo 201-8511, Japan\\
$^{2}$Department of Physics, Science University of Tokyo, 
Tokyo 162-8601, Japan} T. Murayama,$^{1,2 }$ and S. Ono$^{1}$}

\runninghead{Y. Ando, T. Murayama, and S. Ono}
{Magnetotransport Properties of Bi$_{2}$Sr$_{2-x}$La$_x$CuO$_{6}$}
       
\begin{document}

\begin{abstract}
We report that it is possible to obtain a series of high-quality 
crystals of Bi$_{2}$Sr$_{2-x}$La$_x$CuO$_{6}$,
of which the transport properties have been believed to be 
``dirtier" than those of other cuprates.
In our crystals, the normal-state transport properties 
display behaviors which are in good accord with other cuprates; 
for example, in the underdoped region the in-plane resistivity $\rho_{ab}$ 
shows the pseudogap feature and in the overdoped region the $T$ dependence 
of $\rho_{ab}$ changes to $T^n$ with $n>$1.
The characteristic temperatures of the pseudogap deduced from the 
resistivity and the Hall coefficient data are presented.

\end{abstract}

\maketitle

\vspace{0.3in}

A useful way to elucidate the origin of the peculiar normal-state 
properties of high-$T_c$ cuprates is to study their systematic 
evolution upon changing the carrier concentration.
Bi$_{2}$Sr$_{2}$CuO$_{6}$ (Bi-2201) system is an attractive 
candidate for such studies, because the carrier concentration can be 
widely changed by partially replacing Sr with La (to underdope) or Bi 
with Pb (to overdope) \cite{Maeda}.  Moreover, this system allows us 
to study the normal-state in a wider temperature range, because 
the optimum $T_c$ (achieved in Bi$_{2}$Sr$_{2-x}$La$_x$CuO$_{6}$ with 
$x$$\simeq$0.4 \cite{Maeda,Yoshizaki}) is about 30 K, which is lower 
than the optimum $T_c$ of La$_{2-x}$Sr$_{x}$CuO$_{4}$. 
However, a number of problems have been known so far for
Bi-2201 crystals: 
(i) the transport properties of Bi-2201
are quite non-reproducible even among crystals of nominally the 
same composition \cite{Mackenzie,Ando};
(ii) the residual resistivity of $\rho_{ab}$
is usually large (the smallest value reported to date is 
70 $\mu \Omega$cm \cite{Ando,Martin}), as opposed to other systems 
where the residual resistivity in high-quality crystals is negligibly 
small; and 
(iii) the temperature dependence of the Hall coefficient $R_H$ is weak 
and thus the cotangent of the Hall angle $\theta_H$ does not obey the 
$T^2$ law \cite{Mackenzie}, while $\cot \theta_H$$\sim$$T^2$ has 
been almost universally observed in other cuprates.

\begin{figure}
\centerline{\psfig{file=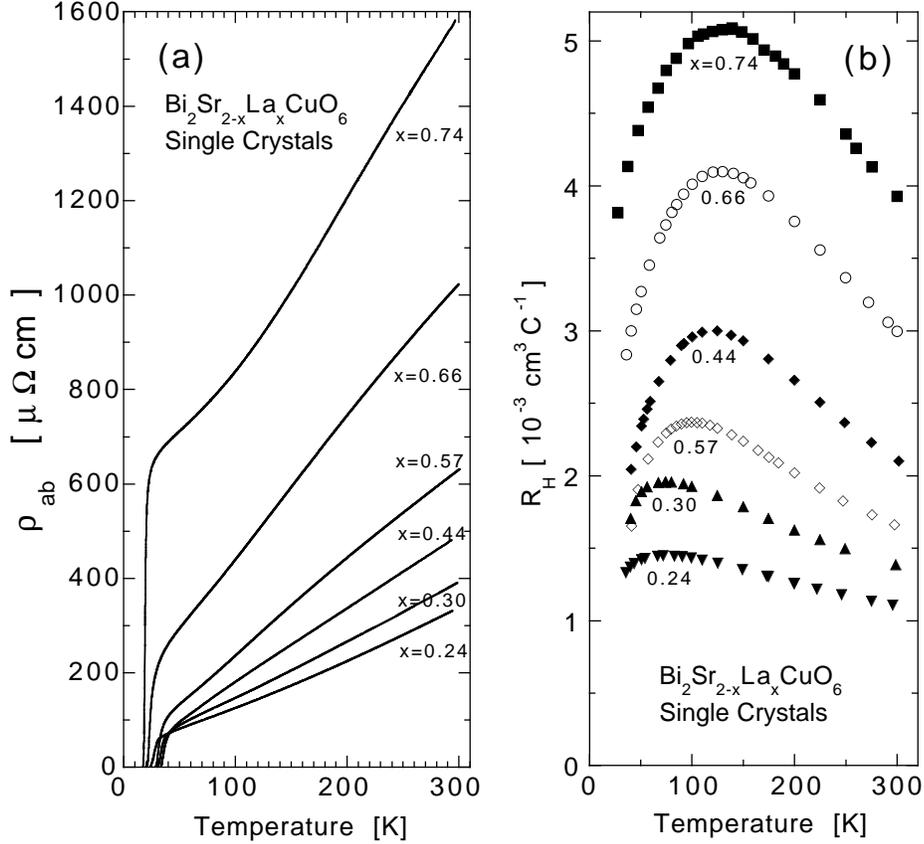,width=12.5cm}}
\caption{$T$ dependence of (a) $\rho_{ab}$ and (b) $R_H$ of the 
BSLCO crystals with various $x$. Note that the extrapolated
residual resistivity of $x$=0.44 sample is 25 $\mu \Omega$cm, which
is the smallest value to date for Bi-2201 or BSLCO.}  
\label{fig1} 
\end{figure}

Here we report the transport properties of a series 
of high-quality Bi-2201 crystals, in which the normal-state 
transport properties display behaviors
that are in good accord with other cuprates.
We show data on $\rho_{ab}(T)$ and $R_H(T)$ for a wide range
of carrier concentrations, from which we extract the characteristic 
temperatures for the pseudogap.

The single crystals of Bi$_{2}$Sr$_{2-x}$La$_x$CuO$_{6}$ (BSLCO) 
are grown using a floating-zone technique in 1 atm of 
flowing oxygen.
Note that pure Bi-2201 is an overdoped system \cite{Maeda}, and 
increasing La doping brings the system from overdoped region 
to underdoped region.
The actual La concentrations in the crystals are 
determined with the ICP analysis.
Here we report crystals with $x$=0.24, 0.30, 0.44, 0.57, 0.66, and 0.74,
for which the zero-resistance $T_c$ is 24, 30, 33.3, 29.2, 21.4 K, 
and 17.3 K, respectively.  
The optimum doping is achieved with $x$$\simeq$0.4, which is consistent 
with previous reports on BSLCO \cite{Maeda,Yoshizaki}.
The optimum zero-resistance $T_c$ of 33 K (which is in agreement with the
Meissner-onset $T_c$) is, to our knowledge, the highest
value ever reported for Bi-2201 or BSLCO system.

\begin{figure}
\centerline{\psfig{file=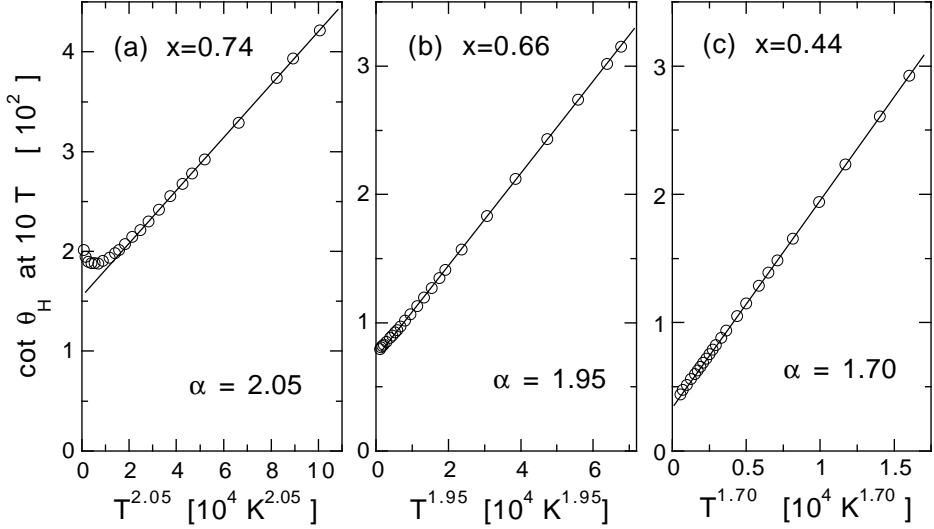,width=12.5cm}}
\caption{Plots of $\cot \theta_H$ vs $T^{\alpha}$ for 
(a) $x$=0.74, (b) $x$=0.66, and (c) $x$=0.44.}  
\label{fig2} 
\end{figure}

Figure 1(a) shows the $T$ dependence of $\rho_{ab}$ for the
six $x$ values in zero field.
Clearly, both the magnitude of $\rho_{ab}$ and its slope  
show a systematic decrease with increasing carrier concentration
(decreasing $x$).
One can see that it is only at the optimum doping that $\rho_{ab}$
shows a good $T$-linear behavior: In the underdoped region, 
$\rho_{ab}(T)$ shows a downward deviation from the $T$-linear 
behavior, which has been discussed to mark the pseudogap
\cite{Ito}. In the
overdoped region, $\rho_{ab}(T)$ shows an upward curvature in the whole
temperature range; the $T$ dependence of $\rho_{ab}$ in the
overdoped region can be well described by $\rho_{ab}$=$\rho_0+AT^n$ 
(with $n$=1.14 and 1.27 for $x$=0.30 and 0.24, respectively),
which is a behavior known to be peculiar for the overdoped cuprates
\cite{Kubo,Takagi}.
Shown in Fig. 1(b) is the $T$ dependence of $R_H$ for the six
samples.  Here again, a clear evolution of $R_H$ with $x$ is observed; 
the change in the magnitude of $R_H$ at 300 K suggests that the carrier concentration is actually reduced roughly by a factor of 4 upon 
increasing $x$ from 0.24 to 0.74.  Note that the $T$ dependence of
$R_H$ is stronger than those previously reported 
\cite{Mackenzie,Ando2}.  In our data, a pronounced peak in $R_H(T)$ 
is clearly observed for all carrier concentrations and the 
position of the peak shifts systematically to higher temperatures 
as the carrier concentration is reduced.

We observed that the cotangent of the Hall angle, $\cot \theta_H$, obeys
a power-law temperature dependence, $T^{\alpha}$, where $\alpha$ is
nearly 2 in underdoped samples ($x$=0.74 and 0.66)
but shows a systematic decrease with increasing carrier concentration
\cite{Murayama}.
Figure 2 shows the plots of $\cot \theta_H$ vs $T^{\alpha}$ for the two
underdoped samples ($x$=0.74 and 0.66) and the optimally-doped 
sample ($x$=0.44).  We note that the $T^2$ law of $\cot \theta_H$
is confirmed for the first time for Bi-2201 in our crystals.
A particularly intriguing fact here is that $\cot \theta_H$ of the 
optimally-doped sample changes as $T^{1.7}$, not as $T^2$,
while $\rho_{ab}$ shows a good $T$-linear behavior.
This suggests that the Fermi-liquid-like behavior of the Hall scattering
rate, $\tau_H^{-1}$$\sim$$T^2$,  
may {\it not} be a generic feature of the optimally-doped cuprates.
The upward deviation from the $T^2$ behavior evident in Fig. 2(a)
for the most underdoped sample ($x$=0.74) is likely to be related to the opening of 
the pseudogap \cite{Xu,Abe}.

\begin{figure}
\centerline{\psfig{file=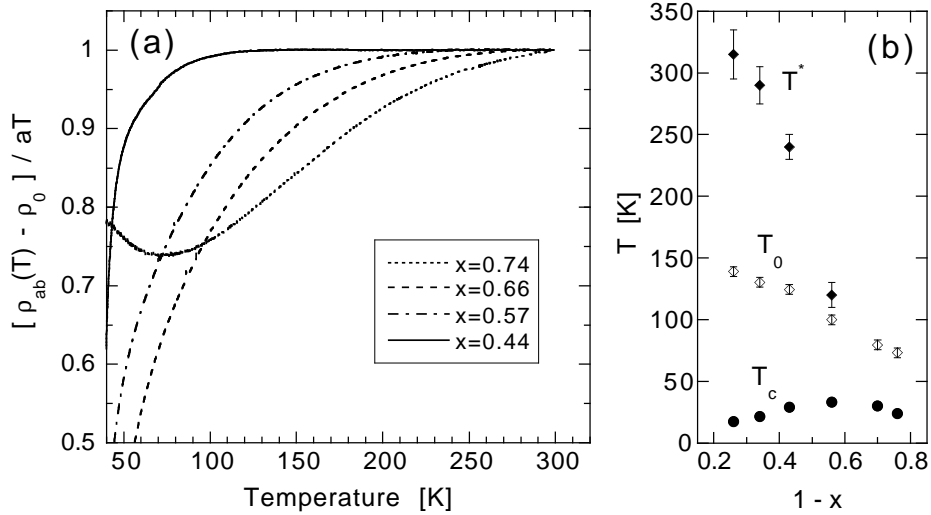,width=12.5cm}}
\caption{(a) Plots of $(\rho_{ab}(T)-\rho_0)/aT$ vs $T$ for the 
optimally-doped and underdoped samples. (b) Phase diagram to
show the two characteristic temperatures for the pseudogap, 
$T^*$ and $T_0$, and the superconducting transition temperature $T_c$.
$T_0$ is defined by the peak in $R_H(T)$.}  
\label{fig3} 
\end{figure}

Due to the limited space, we will concentrate below on the implication 
of our data to the pseudogap in Bi-2201.
As we noted above, the downward deviation from the $T$-linear
behavior in $\rho_{ab}(T)$ has been associated with the pseudogap
and the onset of the deviation at $T^*$ gives a characteristic 
temperature for the pseudogap \cite{Ito}.
Figure 3(a) shows the plot which emphasizes the deviation from the 
$T$-linear behavior to determine $T^*$.  We should mention that 
this type of plot is subject to some arbitrariness and thus the 
errors in $T^*$ are inevitably large.  
(Interestingly, in Fig. 3(a), even the optimally-doped sample shows a 
$T^*$ which is well above $T_c$.)
It has recently been recognized that there are two different 
characteristic temperatures for the pseudogap \cite{Batlogg}, 
and $T^*$ corresponds to the higher characteristic temperature 
for the pseudogap.
Also, it was proposed very recently that the peak in the $T$ dependence 
of $R_H$ may mark the lower characteristic temperature $T_0$ for the 
pseudogap \cite{Xu,Abe}, as does the NMR relaxation rate, ARPES, or the
tunnelling spectroscopy \cite{Timusk}.  
In our samples, the peak in $R_H(T)$ moves to higher temperatures
as the carrier concentration is reduced, which is consistent with the
behavior of the pseudogap.
We note, however, that it is not clear whether the peak in $R_H(T)$ 
observed in the {\it overdoped} samples really corresponds to the 
pseudogap.
For a more detailed discussion on the relation 
between the peak in $R_H(T)$ and the pseudogap, please refer to Ref. 
[12]. 

Figure 3(b) shows the phase diagram, $T$ vs $1-x$, for our BSLCO.
Note that the horizontal axis is taken to be $1-x$ for convenience;
this way, the left hand side of the plot corresponds to underdoping.
One can see in Fig. 3(b) that a significant portion of
the phase diagram has been covered 
and all the plotted temperatures, $T^*$, $T_0$,
and $T_c$, show good systematics with the carrier concentration.
It is intriguing that the magnitudes of the 
characteristic temperatures for the pseudogap is quite similar to
other cuprates \cite{Timusk}, while the $T_c$ of the present system
is the lowest among the major cuprates.

In summary, we present the data of the in-plane resistivity,
Hall coefficient, and the Hall angle of a series of high-quality 
La-doped Bi-2201 crystals in a wide range of carrier concentrations.
The normal-state transport properties of our Bi-2201 crystals
show systematics that can be considered to be ``canonical" to cuprates.
The characteristic temperatures for the pseudogap were deduced 
from the data to construct a phase diagram.


\begin{thebibliography}{9}

\bibitem{Maeda} 
A. Maeda {\it et al.},
{\it Phys. Rev. B} {\bf 41}, 6418 (1990).

\bibitem{Yoshizaki} 
R. Yoshizaki {\it et al.}, {\it Physica C} {\bf 224}, 121 (1994).

\bibitem{Mackenzie}
A. P. Mackenzie {\it et al.}, {\it Phys. Rev. B} {\bf 45}, 527 (1992).

\bibitem{Ando}
Y. Ando {\it et al.},
{\it Phys. Rev. Lett.} {\bf 77}, 2065 (1996); {\bf 79}, 2595(E) (1997).

\bibitem{Martin}
S. Martin {\it et al.}, {\it Phys. Rev. B} {\bf 41}, 846 (1990).

\bibitem{Ito}
T. Ito, K. Takenaka, and S. Uchida, {\it Phys. Rev. Lett.} {\bf 70}, 3995 (1993).

\bibitem{Kubo}
Y. Kubo {\it et al.},
{\it Phys. Rev. B} {\bf 43}, 7875 (1991).

\bibitem{Takagi}
H. Takagi {\it et al.}, {\it Phys. Rev. Lett.} {\bf 69}, 2975 (1992).

\bibitem{Ando2}
Y. Ando {\it et al.}, {\it Phys. Rev. B} {\bf 56}, R8530 (1997).

\bibitem{Murayama}
Y. Ando and T. Murayama, preprint (cond-mat/9812334).

\bibitem{Xu}
Z. A. Xu. Y. Zhang, and N. P. Ong, preprint (cond-mat/9903123).

\bibitem{Abe}
Y. Abe, K. Segawa, and Y. Ando, preprint (cond-mat/9905274).

\bibitem{Batlogg}
B. Batlogg and V. J. Emery, {\it Nature} {\bf 382}, 20 (1996).

\bibitem{Timusk}
For a review, see T. Timusk and B. Statt, preprint (cond-mat/9905219).

\end{thebibliography}
\end{document}